\documentclass{PoS}
\title{INFN Camera demonstrator for the Cherenkov Telescope Array}
\ShortTitle{INFN Camera demonstrator for CTA}
\author{ G. Ambrosi,$^a$ M. Ambrosio,$^b$ C. Aramo,$^b$ B. Bertucci,$^{ac}$ E. Bissaldi,$^{o}$ M. Bitossi,$^f$ 
S. Brasolin,$^g$ 
G. Busetto,$^l$ R. Carosi,$^h$ S. Catalanotti,$^{bi}$ M.A. Ciocci,$^{hj}$ R. Consoletti,$^r$ P. Da Vela,$^{hj}$ 
F. Dazzi,$^k$ A. De Angelis,$^l$\thanks{email:Alessandro.DeAngelis@pd.infn.it, Nicola.Giglietto@ba.infn.it, Carlo.Vigorito@to.infn.it}$~$ B. De Lotto,$^{ny}$ 
F. de Palma,$^{mo}$ R. Desiante,$^{gn}$ 
T. Di Girolamo,$^{bi}$ C. Di Giulio,$^p$ M. Doro,$^{kl}$ D. D'Urso,$^{aw}$ G. Ferraro,$^{or}$ F. Ferrarotto,$^s$ 
F. Gargano,$^r$ N. Giglietto,$^{or}$ F. Giordano,$^{or}$ G. Giraudo,$^g$ M. Iacovacci,$^{bi}$ M. Ionica,$^a$ M. Iori,$^{st}$ 
F. Longo,$^{de}$ M. Mariotti,$^{lq}$, S. Mastroianni,$^b$ M. Minuti,$^h$ A. Morselli,$^p$ R. Paoletti,$^{hj}$ 
G. Pauletta,$^{ny}$ R. Rando,$^{lq}$ G. Rodriguez Fernandez,$^p$
A. Rugliancich,$^{hj}$ D. Simone,$^{o}$ C. Stella,$^{dn}$ A. Tonachini,$^{gu}$ P. Vallania,$^{gv}$ L. Valore,$^{bi}$, V. Vagelli,$^a$  
V. Verzi,$^p$ and \speaker{C. Vigorito},$^{gu}$ for the CTA Consortium\thanks{Full consortium author list at
http://cta-observatory.org}\\
\llap{$^a$} INFN-Sezione di Perugia, Italy\\
\llap{$^b$} INFN-Sezione di Napoli, Italy\\
\llap{$^c$} Universit\`a di Perugia, Italy\\
\llap{$^d$} INFN-Sezione di Trieste, Italy\\
\llap{$^e$} Dipartimento di Fisica, Universit\`a di Trieste, Italy\\
\llap{$^f$} European Gravitational Observatory (EGO), Cascina (PI), Italy\\
\llap{$^g$} INFN-Sezione di Torino, Italy\\
\llap{$^h$} INFN-Sezione di Pisa, Italy\\
\llap{$^i$} Universit\`a di Napoli "Federico II", Italy\\
\llap{$^j$} Universit\`a di Siena, Italy\\
\llap{$^k$} Max-Planck-Institut f\"ur Physik, M\"unchen, Germany\\
\llap{$^l$} INFN-Sezione di Padova, Italy\\
\llap{$^m$} Universit\`a Telematica Pegaso, Napoli, Italy\\
\llap{$^n$} Dipartimento di Chimica, Fisica e Ambiente, Universit\`a di Udine, Italy\\
\llap{$^o$} INFN-Sezione di Bari, Italy\\
\llap{$^p$} INFN-Sezione di Roma "Tor Vergata", Italy\\
\llap{$^q$} Universit\`a di Padova, Italy\\
\llap{$^r$} Dipartimento inter-ateneo di Fisica dell'Universit\`a e del Politecnico di Bari, Italy\\
\llap{$^s$} INFN-Sezione di Roma, Italy\\
\llap{$^t$} Universit\`a di Roma "La Sapienza", Italy\\
\llap{$^u$} Dipartimento di Fisica, Universit\`a di Torino, Italy\\
\llap{$^v$} INAF, Osservatorio Astrofisico di Torino, Italy\\
\llap{$^w$} ASI Science Data Center (ASDC), Roma, Italy\\
\llap{$^y$} INFN-Sezione di Trieste, Gruppo collegato di Udine, Italy\\\\
}

\abstract{The Cherenkov Telescope Array is a world-wide project for a new generation of ground-based Cherenkov
telescopes of the Imaging class with the aim of exploring the highest
energy region of the electromagnetic spectrum. With two planned arrays, one for each hemisphere, it will
guarantee a good sky coverage in the energy range from a few tens of GeV to hundreds of TeV, with
improved angular resolution and a sensitivity in the TeV energy region better by one order of
magnitude  than  the  currently  operating  arrays. In  order  to  cover  this  wide  energy  range,  three
different telescope types are envisaged, with different mirror sizes and focal plane features.
In particular, for the highest energies  a possible design is  a dual-mirror Schwarzschild-Couder
optical scheme, with a compact focal plane. A silicon photomultiplier (SiPM) based camera is being
proposed as a solution to match the dimensions of the pixel (angular size of ~ 0.17 degrees).
INFN is developing a camera demonstrator made by 9 Photo Sensor Modules (PSMs, 64 pixels
each, with total coverage 1/4 of the focal plane) equipped with FBK (Fondazione Bruno Kessler,
Italy) Near UltraViolet High Fill factor SiPMs and Front-End Electronics (FEE) based on a Target 7 ASIC,  a  16  channels   fast  sampler   (up  to  2GS/s)  with  deep  buffer,   self-trigger   and  on-demand
digitization capabilities specifically developed for this purpose. The pixel dimensions of $6\times6$ mm$^2$
lead to a very compact design with challenging problems of thermal dissipation.
A modular structure, made by copper frames hosting one PSM and the corresponding FEE, has been conceived, with a water cooling system to keep the required
working temperature. The actual design, the adopted technical solutions and the achieved results for
this demonstrator are presented and discussed.}

\FullConference{The 34th International Cosmic Ray Conference,\\
		30 July- 6 August, 2015\\
		The Hague, The Netherlands}

\begin{document}

\section{Introduction}
The Cherenkov Telescope Array (CTA) is a world-wide project dedicated to
ground-based gamma-ray astronomy, with the participation of the majority of EU countries \cite{CTA1,CTA2}.
The CTA experiment is planned to include two arrays of more than 100 telescopes in total which will detect gamma rays from few tens 
of GeV to more than 100 TeV. In the core energy region (1-10) TeV it will improve by at least one order of magnitude the sensitivity of the Very High Energy (VHE) 
telescopes currently in operation (e.g. H.E.S.S. \cite{hess1,hess2}, MAGIC \cite{magic1,magic2} and VERITAS \cite{veritas}) 
and will be a facility open to the whole astrophysical community.\\ 
The success of such a complex project relies on challenging scientific and technical developments, four of which could benefit 
from Italian Research Centres know-how and innovative approach: (1) the industrial production of low-cost telescopes ideally suited to detect
 Cherenkov emission of relativistic particles produced by celestial gamma rays; (2) the development and production of 
silicon-based photon detectors; (3) the development of novel front-end electronics to digitize, read and transfer the detector signals; 
and (4) the extremely rapid analysis pipeline based on the use of  Graphical Processing Units (GPUs), in order  to make it possible 
to perform real time analysis.

\begin{figure}[!ht]
\vspace{0.5cm}
\begin{center}
\includegraphics[width=.85\linewidth]{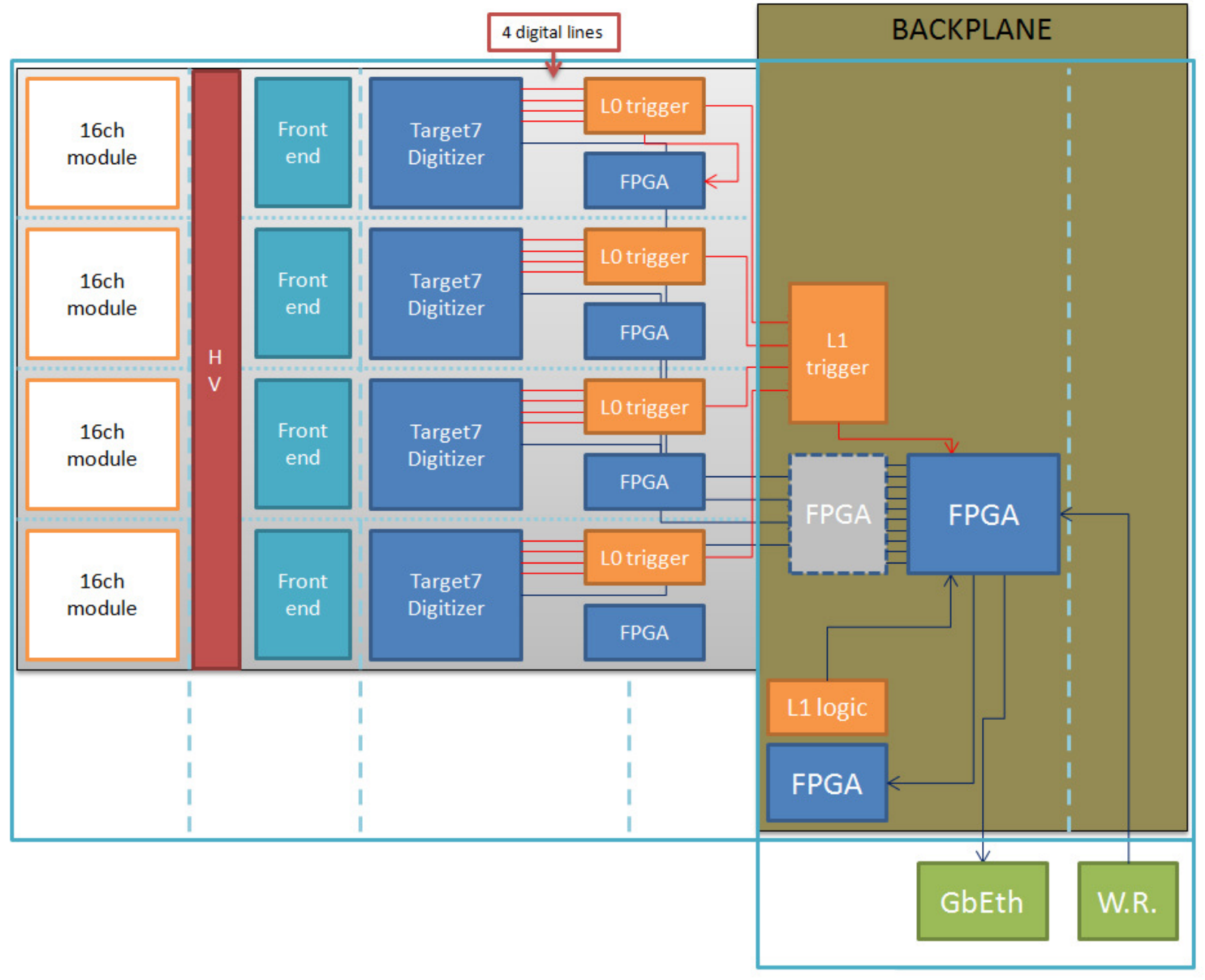}
\caption{Sketch of the readout system of a PSM.}
\label{Fig_4}
\end{center}
\end{figure}

The {\it Istituto Nazionale di Fisica Nucleare} (INFN) and the {\it Istituto Nazionale di Astrofisica} (INAF) groups, together 
with some industrial partners \cite{caen,sitael}, joined to create the Italian excellence program "Progetto Premiale TECHE.it" in 2014 for the construction of a SiPM-based camera for CTA 
telescopes.  The aim of this project is to push the features of SiPM sensors to fit CTA telescopes requirements and
 at the same time to produce a realistic demonstrator, in which most of the items necessary to design a SiPM based camera, 
including electronics, mechanical design and the heat diffusion aspects, are well evaluated. This project was driven  together with
the industrial partners to the construction of a complete prototype.
The demonstrator will consist of a SiPM plane made of  $6\times6$ mm$^2$ pixels.
The signals from the silicon photomultipliers channels will be conditioned by a PCB housing the electronics, which will be coupled to a board dedicated to the  trigger formation and digitization of the voltage signals from the preamplifiers.
In this paper we summarize the state of development of the camera demonstrator. Sections \ref{fp},\ref{fee},\ref{ms} show the results
obtained in the development of the focal plane, front-end electronics and mechanics of the prototype respectively.

\section{Focal plane and Sensors}
\label{fp}
\begin{figure}[!ht]
\vspace{0.5cm}
\begin{center}
\includegraphics[width=.65\linewidth]{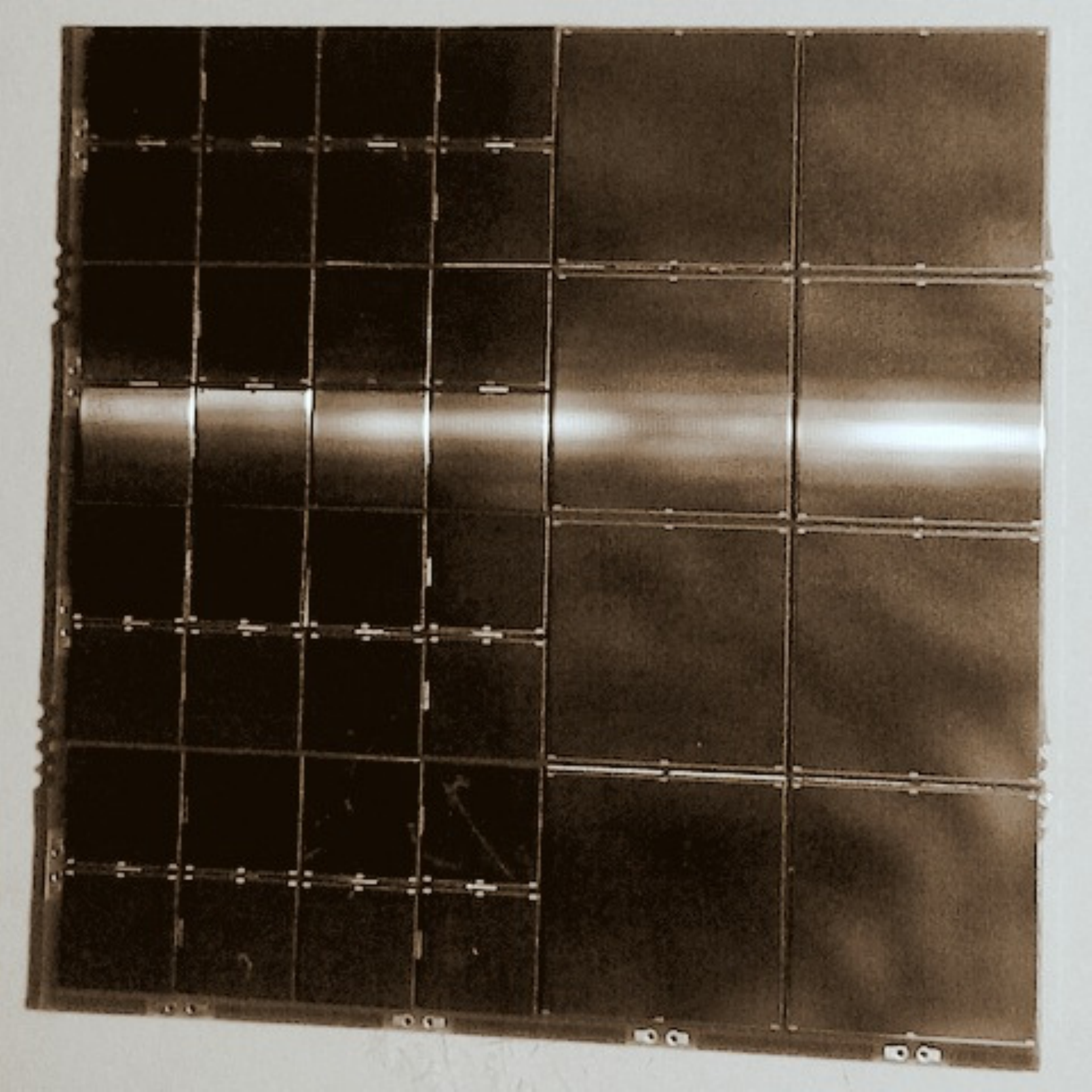}
\caption{Sensors matrix of 1/4 of a PSM. Two different sizes for SiPM units have been included in this prototype in order to verify
their different perfomances. SiPMs are produced by Fondazione Bruno Kessler (Trento, Italy) \cite{fbk}.}
\label{Fig_1}
\end{center}
\end{figure}

The focal plane corresponding to a Photo Sensor Module (PSM) will consist of an array of $8\times8$ pixels, each one of 
$6\times6$ mm$^2$ area and 0.17 degrees of Field Of View (FOV). Fig. \ref{Fig_1} shows a picture of  1/4 of a PSM,  with 
two different pixel configurations and different
sensor sizes and arrangements, namely: 
8 units of $6\times6$ mm$^2$  sensors on the right side,
and 32 units of $3\times3$ mm$^2$ sensors on the left side, summed up
in blocks of four to obtain the equivalent area of $6\times6$ mm$^2$  for each pixel.
This was done to verify and
compare the performance of the two configurations based on different SiPM units.
The smaller sensors, already fully tested (see \cite{scineghe1} \cite{scineghe2} for reference), 
represent a possible backup solution in case of production failure
of the larger sensors which are the best geometrical solution for the requested 
area and FOV.\\
Although the solution of having 
monolithic $6\times6$ mm$^2$ pixels SiPMs would be preferable,
smaller sensors have the advantage of reduced 
costs due to higher production yield. However, 
their difference in breakdown voltage and gain
may result in a worse single photo-electron (p.e.) energy resolution
compared to the larger sensors.

The voltage to the SiPMs is biased
by means of pads outside the active area and all SiPMs 
are bonded in daisy-chain.
This solution has been adopted to optimize
the distance between adjacent pixels,
obtaining a fill factor $>80$\%.
With this design, a negative HV is supplied
to the SiPMs to deplete the junction, while
the cathode is directly connected to the 
amplifier's negative input and grounded
through the virtual short (see next paragraph).
\section{Front-End Electronics}
\label{fee}
\subsection{Preamplifier}
The current signal generated by photoconversion
in SiPMs is fed to the negative input
of an AD8000 operational amplifier.
The basic scheme is that of a trans-impedance
amplifier having a 20 $\Omega$ input resistance and 
a 1 k$\Omega$ feedback resistance.
A Pole-Zero cancellation network \cite{GOL13} was added 
to the preamplifier design to eliminate the long tail
of the current signal. A 470 pF capacitance was placed in
parallel to a 1 k$\Omega$ resistance and closed over
a 50 $\Omega$ load. 

\subsection{The Readout System}
The readout system is based on a modular approach (see Fig. \ref{Fig_4}) 
where each module is interfaced to a group of 16 pixels of the PSM. 
Each readout board hosts the front-end pre-amplifiers that condition the signal for the Target 7 (T7) ASIC, responsible 
for the sampling and digitization of the input signals. The T7 is configured 
by a local FPGA (Altera Cyclone IV 
model EP4CE22E144) that is also responsible for the data management.
Below the PSM an interface board hosts the high voltage generation that is controlled by one local FPGA. 
In a future version of the module readout system there will be one single FPGA responsible for the T7 operation 
and HV generation and control. The original HV generation is based on a programmable bias supply chip (Maxim MAX1932) that
 can provide up to 90 Volt output.
A feature of the T7 ASIC chip is that it generates four digital trigger lines, each one produced by on-chip
 comparators on analog sum of four pixels group. These digital lines are used to produce a local trigger signal,
 called level 0 (L0), and the L0 information is transferred to a backplane FPGA (Altera Cyclone III model EP3C55F484)
 that combines them to generate a higher level trigger, called level 1 (L1). The L1 triggers from several modules can
 be combined to produce a camera trigger that enables the data transmission from the local to the backplane FPGA and 
to the data acquisition system via a Gbit ethernet connection based on the Marvell 88E1111 transceiver.

\section{The Mechanical Structure: thermal study}
\label{ms}

\begin{figure}[!ht]
\vspace{0.5cm}
\begin{center}
\includegraphics[width=.85\linewidth]{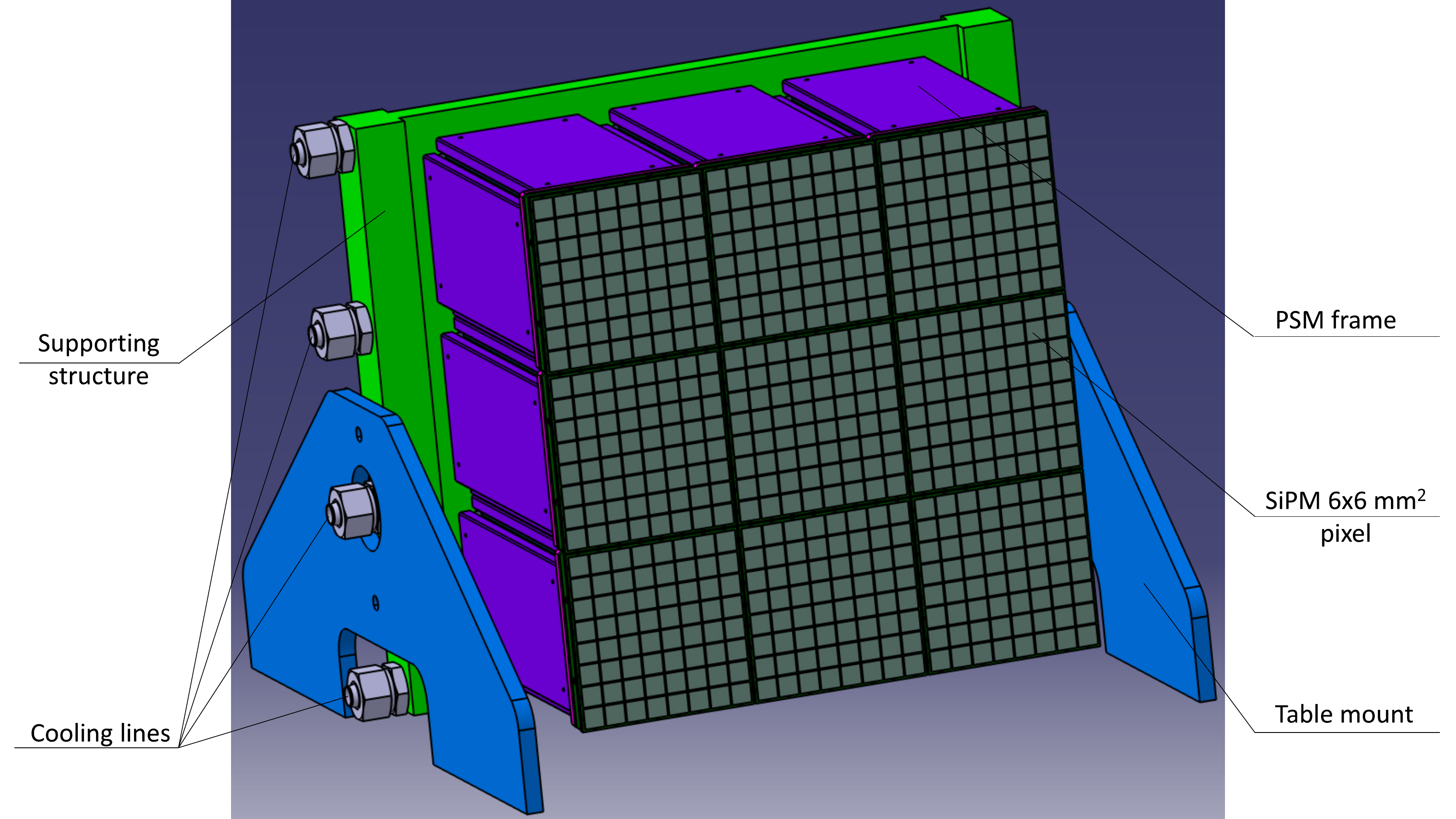}
\caption{Pictorial view of the mechanical structure of the demonstrator.}
\label{Fig_5}
\end{center}
\end{figure}

For the construction of the mechanical structure  able to host the PCBs and the focal plane SiPMs of a matrix of $3\times3$ PSMs we de%
cided to start with a flat configuration, even if the focal plane of a Schwarzschild Couder optical scheme is curved. This allowed us to focus our efforts on the mechanical and thermal problems instead of geometry.
Due to the very compact design with a focal plane fully covered by the SiPM detectors, the main issue is 
the thermal dissipation.
Each PSM is equipped with 4 PCBs carrying the FEE for 16 channels each.
The thermal dissipation is mainly due to the AD8000 operational preamplifiers (16x13.5 mW) 
and the Target 7 ASIC and FPGA (100 mW each) giving a total power consumption of 0.416 W for each PCB and 1.664 W for the whole PSM frame.
The total  power for the demonstrator is 15 W. This value seems quite low but being concentrated in an object of dimensions 16x16 cm$^2$ it makes a challenging problem to keep the electronics at a reasonable temperature.
At this purpose, two very different solutions were envisaged to host the PCBs inside the PSM frame: in the first one the 4 PCBs were located in parallel into the module allowing the air circulation by a fan; in the second one the PCBs were on the contrary located on the sides of the PSM and the heat was carried out from the electronic components to the supporting structure by means of thermal pads and copper frames.
After some preliminary calculations on both configurations, the second one was selected provided that an efficient thermal bridge is made carrying out the heat from the PCB to the supporting structure. 
The latter is then cooled down by means of water flowing through its main arms and chilled by an external device.
Starting from this general scheme the project has been refined obtaining the design that is currently under construction (see Fig. \ref{Fig_5}): 
first of all the chosen material for the realization of the whole mechanical structure is copper, thanks to the optimum thermal conductivity (k=450 W/m$\cdot$K compared to K=150 W/m$\cdot$K for aluminium); moreover each PCB is covered by a thermal pad (i.e. THERM-A-GAP by Parker Chomerics \cite{parker}) with a dissipation power k=6.5 W/m$\cdot$K and thermal grease GEL30G or GEL652 with k=3-3.5 W/m$\cdot$K.
The thermal behaviour of the entire structure has been accurately simulated obtaining a maximum temperature of 31.7$^{\circ}$C with a water flow of $2$ l/min at 18$^{\circ}$C on each of the 4 independent cooling lines. 
This relatively high temperature should avoid the dangerous formation of dew on the FEE and the focal plane.
The PSM frame allows also the arrangment of the SiPM HV distribution located on the bottom of a square PCB mounted on the top frame.

Since these PCBs must have a minimum separation to avoid focal plane dead areas (in our case $< 1$ mm), the resulting assembly is very tight but due to its modularity it allows the replacement of an entire PSM frame in a quite easy way, also considering that the water cooling affects the supporting structure only.
The demonstrator mechanical structure is currently under construction at the Technological Laboratory of INFN in Turin, Italy.
When completed, it will be equipped with dummy PCBs with suitable resistors to obtain the right power dissipation at the expected location, and a matrix of temperature sensors to monitor the heat flow.
The test will be done in a climatic chamber with different temperature and humidity conditions and with the structure connected to a chiller able to supply water at the requested flow and temperature.
In this way we will verify the thermal simulation results and optimize 
the cooling parameters changing the working conditions over the range set by the CTA requirements (Observation Temperature: -15 to +25$^{\circ}$C; Observation Humidity: 2 to 90\%).\\

\begin{figure}[!ht]
\vspace{0.5cm}
\begin{center}
\includegraphics[width=.85\linewidth]{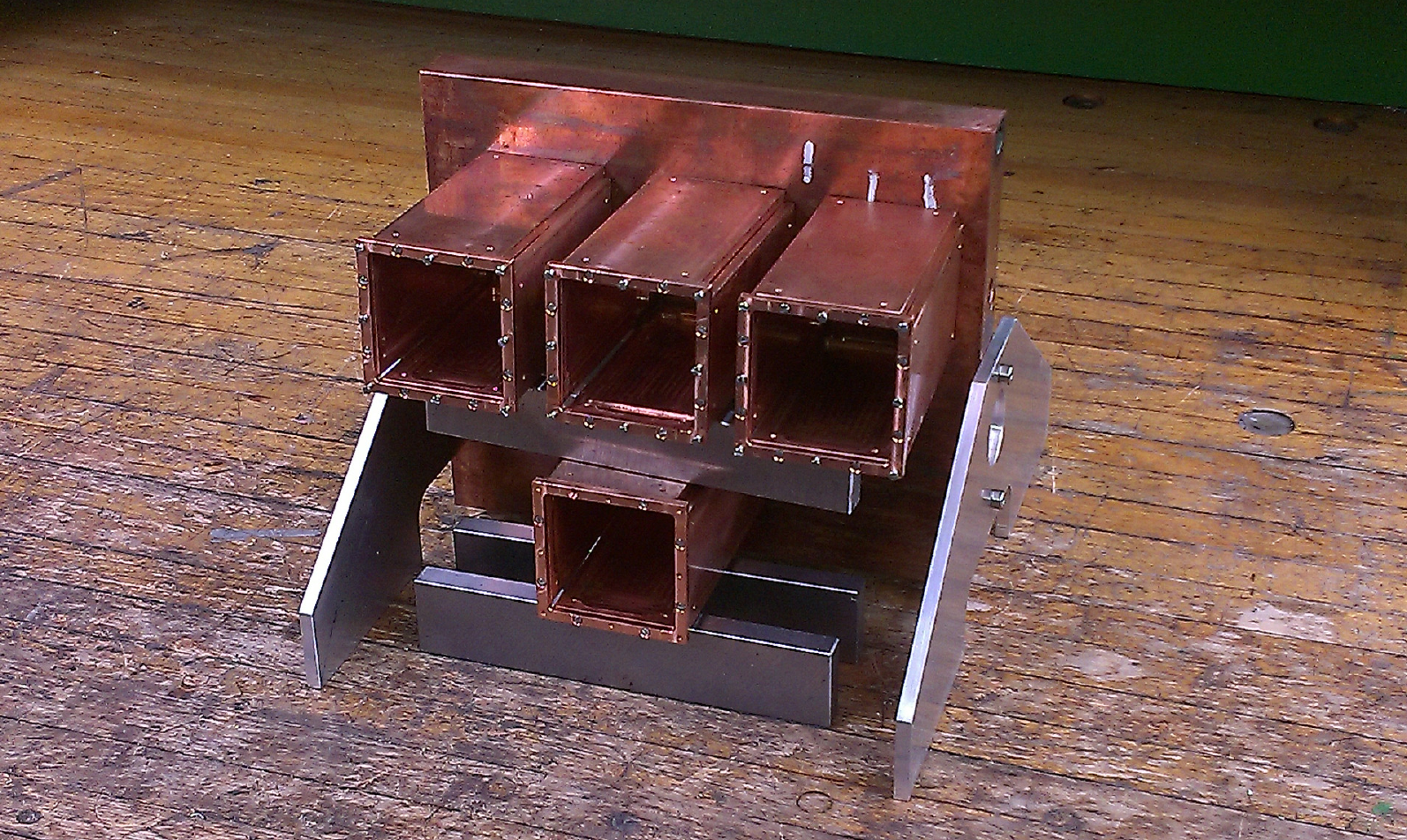}
\caption{Picture of the mechanical structure of the demonstrator at INFN-Torino: status at June 2015.}
\label{Fig_6}
\end{center}
\end{figure}

A picture of the mechanical structure, while in construction at INFN-Torino, is shown in Fig. \ref{Fig_6}. 

We expect to start the laboratory tests late in July, 2015, therefore an update on the first results will be given at the Conference. 

\section{Conclusions}
In the frame of "Progetto Premiale TECHE.it", a joint venture between INFN, INAF and industrial partners, a camera demonstrator
of the SST telescope, made by 9 PSMs, has been designed and is currently under construction. 
This paper discussed the state of the art  achieved in the 
construction of 1/4 of the full camera.  Possible different solutions for the sensors, Front-End 
Electronics and mechanical structure have been investigated. The expected completion of the camera, including data acquisition system, is foreseen by the end of 2015.  

\section{Acknowledgments} 
We gratefully acknowledge support from the agencies and organizations 
listed under Funding Agencies at this website: http://www.cta-observatory.org/.

\end{document}